\documentclass{article}
\usepackage[T1]{fontenc} 
\usepackage[utf8]{inputenc} 
\usepackage{ismir, amsmath,cite,url, amssymb}
\usepackage{graphicx}

\usepackage{color}
\usepackage{booktabs}
\usepackage[bookmarks=false]{hyperref}

\usepackage{lineno}

\title{DrumGAN: Synthesis of Drum Sounds with Timbral Feature Conditioning Using Generative Adversarial Networks}

\threeauthors
  {Javier Nistal} {Sony CSL\\ Paris, France}
  {Stefan Lattner} {Sony CSL\\Paris, France}
  {Ga\"{e}l Richard} {LTCI, T\'{e}l\'{e}com Paris \\Institut Polytechnique de Paris, France}

\begin{document}

\maketitle
\begin{abstract}
Synthetic creation of drum sounds (e.g., in drum machines) is commonly performed using analog or digital synthesis, allowing a musician to sculpt the desired timbre modifying various parameters. Typically, such parameters control low-level features of the sound and often have no musical meaning or perceptual correspondence. With the rise of Deep Learning, data-driven processing of audio emerges as an alternative to traditional signal processing. This new paradigm allows controlling the synthesis process through learned high-level features or by conditioning a model on musically relevant information. In this paper, we apply a Generative Adversarial Network to the task of audio synthesis of drum sounds. By conditioning the model on perceptual features computed with a publicly available feature-extractor, intuitive control is gained over the generation process. The experiments are carried out on a large collection of kick, snare, and cymbal sounds. We show that, compared to a specific prior work based on a U-Net architecture, our approach considerably improves the quality of the generated drum samples, and that the conditional input indeed shapes the perceptual characteristics of the sounds. Also, we provide audio examples and release the code used in our experiments.\footnote{\url{https://github.com/SonyCSLParis/DrumGAN}}
\end{abstract}

\section{introduction}
\label{sec:introduction}

Drum machines are electronic musical instruments that create percussion sounds and allow to arrange them in patterns over time. The sounds produced by some of these machines are created synthetically using analog or digital signal processing. For example, a simple snare drum can be synthesized by generating noise and shaping its amplitude envelope \cite{sos} or, a bass drum, by combining low-frequency harmonic sine waves with dense mid-frequency components \cite{sos1}. The characteristic sound of this synthesis process contributed to the cult status of electronic drum machines in the '80s.


Data-driven processing of audio using Deep Learning (DL) emerged as an alternative to traditional signal processing.  This new paradigm allows us to steer the synthesis process by manipulating learned higher-level latent variables, which provide a more intuitive control compared to conventional drum machines and synthesizers. In addition, as DL models can be trained on arbitrary data, comprehensive control over the generation process can be enabled without limiting the sound characteristic to that of a particular synthesis process. For example, Generative Adversarial Networks (GANs) allow to control drum synthesis through their latent input noise \cite{wavegan} and Variational Autoencoders (VAE) can be used to create variations of existing sounds by manipulating their position in a learned timbral space \cite{planetdrums}. However, an essential issue when learning latent spaces in an unsupervised manner is the missing interpretability of the learned latent dimensions. This can be a disadvantage in music applications, where comprehensible interaction lies at the core of the creative process. Therefore, it is desirable to develop a system which offers expressive and musically meaningful control over its generated output. A way to achieve this, provided that suitable annotations are available, is to feed higher-level conditioning information to the model. The user can then manipulate this conditioning information in the generation process. Along this line, some works on sound synthesis have incorporated pitch-conditioning \cite{wavenetae, gansynth}, or categorical semantic tags \cite{flowsynth}, capturing rather abstract sound characteristics. In the case of drum pattern generation, there are neural-network approaches that can create full drum tracks conditioned on existing musical material \cite{drumnet}. 

In a recent study \cite{aramires}, a U-Net is applied to neural drum sound synthesis, conditioned on continuous perceptual features describing timbre (e.g., boominess, brightness, depth). These features are computed using the Audio Commons timbre models.\footnote{\scriptsize\url{https://github.com/AudioCommons/ac-audio-extractor}} Compared to prior work, this \emph{continuous} feature conditioning (instead of using categorical labels) for audio synthesis provides more fine-grained control to a musician. However, this U-Net approach learns a deterministic mapping of the conditioning input information to the synthesized audio. This limits the model's capacity to capture the variance in the data, resulting in a sound quality that does not seem acceptable in a professional music production scenario.

In this paper, we build upon the same idea of conditional generation using continuous perceptual features, but instead of a U-Net, we employ a Progressive Growing Wasserstein GAN (PGAN) \cite{Karras2017}. Our contribution is two-fold. First, we employ a PGAN on the task of conditional drum sound synthesis. Second, we use an auxiliary regression loss term in the discriminator as a means to control audio generation based on the conditional features. We are not aware of previous work attempting \emph{continuous} sparse conditioning of GANs for musical audio generation. We conduct our experiments on a dataset of a large variety of kick, snare, and cymbal sounds comprising approximately 300k samples. Also, we investigate whether the feature conditioning improves the quality and coherence of the generated audio. For that, we perform an extensive experimental evaluation of our model, both in conditional and unconditional settings. We evaluate our models by comparing the Inception Score (IS), the Fréchet Audio Distance (FAD), and the Kernel Inception Distance (KID). Furthermore, we evaluate the perceptual feature conditioning by testing if changing the value of a specific input feature yields the expected change of the corresponding feature in the generated output. Audio samples of DrumGAN can be found on the accompaniment website (see Section \ref{sec:res}).

The paper is organized as follows: In Section \ref{sec:sota} we review previous work on audio synthesis, and in Section \ref{sec:exp} we describe the experiment setup. Results are presented in Section \ref{sec:res}, and we conclude in Section \ref{sec:con}.

\section{Previous Work}\label{sec:sota}
Deep Generative modeling is a topic that has gained a lot of interest during the last years. This has been possible partly due to the growing amount of large-scale datasets of different modalities \cite{wavenetae, imagenet} coupled with groundbreaking research on generative neural networks \cite{wavenet, vae, Goodfellow2013, Karras2017, pixelrnn}. In addition to the methods listed in the introduction focusing on drums sound generation, a number of other studies have applied deep learning methods to address general audio synthesis. Autoregressive models for raw audio have been very influential in the beginning of this line of research, and still achieve state of the art in different audio synthesis tasks \cite{wavenet, wavenetae, melnet, tacotron2}. Approaches using Variational Auto-Encoders \cite{vae} allow manipulating the audio in latent spaces learnt i) directly from the audio data \cite{planetdrums}, ii) by imposing musically meaningful priors over the structure of these spaces \cite{flowsynth, EslingCB18, Philippe}, or  iii) by restricting such latent codes to discrete representations \cite{jukebox}. GANs have been extensively applied to synthesis of speech \cite{SaitoTS18} and domain adaptation \cite{Kaneko, timbertron} tasks. The first of its kind applying adversarial learning to the synthesis of musical audio is  WaveGAN \cite{wavegan}. This architecture was shown to synthesize audio from a variety of sound sources, including drums, in an unconditional way. Recent improvements in the quality and training stability of GANs \cite{Karras2017, Gulrajani2017, SalimansGZCRCC16} resulted in methods that outperform WaveNet baselines on the task of audio synthesis of musical notes using sparse conditioning labels representing the pitch content \cite{gansynth}. A few other works have used GANs with rather strong conditioning on prior information for tasks like Mel-spectrum inversion \cite{melgan} or audio domain adaptation \cite{Hosseini, MichelsantiT17}. Recently, other promising related research incorporates prior domain-knowledge into the neural network, by embedding differentiable signal processing blocks directly into the architecture \cite{ddsp}.

\section{Experiment Setup}\label{sec:exp}
In this Section details are given about the conducted experiment, including the data used, the model architecture and training details, as well as the metrics employed for evaluation.

\subsection{Data}
In the following, we briefly describe the drum dataset used throughout our experiments, as well as the Audio Commons feature models, with which we extract perceptive features from the dataset.

\subsubsection{Dataset}
For this work, we make use of an internal, non-publicly available dataset of approximately 300k one-shot audio samples aligned and distributed across a balanced set of kick, snare, and cymbal sounds. The samples originally have a sample rate of 44.1kHz and variable lengths. In order to make the task simpler, each sample is shortened to a duration of one second and down-sampled to a sample rate of 16kHz. For each audio sample, we extract perceptual features with the Audio Commons timbre models (see Section \ref{sec:acommons}). We perform an 90\% / 10\% split of the dataset for validation purposes. The model is trained on the real and imaginary components of the Short-Time Fourier Transform (STFT), which has been shown to work well in \cite{nistal1}. We compute the STFT using a window size of 2048 samples and 75\% overlapping. The generated spectrograms are then simply inverted back to the signal domain using the inverse STFT.

\subsubsection{Audio-Commons Timbre Models}\label{sec:acommons}
The Audio Commons project\footnote{\scriptsize\url{https://www.audiocommons.org/2018/07/15/audio-commons-audio-extractor.html}} implements a collection of perceptual models of features that describe high-level timbral properties of the sound. These features are designed from the study of popular timbre ratings given to a collection of sounds obtained from Freesound\footnote{\scriptsize\url{https://freesound.org/}}. The models are built by combining existing low-level features found in the literature (e.g., spectral centroid, dynamic-range, spectral energy ratios, etc), which correlate with the target properties enumerated below. All features are defined in the range [0-100]. We employ these features as conditioning to the generative model. For more information, we direct the reader to the project deliverable.\textsuperscript{3}

\begin{itemize}
    \item \textbf{brightness}: refers to the clarity and amount of high-pitched content in the analyzed sound. It is computed from the spectral centroid and the spectral energy ratio. 
    \vspace*{-0.2cm}
    \item \textbf{hardness}: refers to the stiffness or solid nature of the acoustic source that could have produced a sound. It is estimated using a linear regression model on spectral and temporal features extracted from the attack segment of a sound event.
    \vspace*{-0.2cm}
    \item \textbf{depth}: refers to the sensation of perceiving a sound coming from an acoustic source beneath the surface. A linear regression model estimates depth from the spectral centroid of the lower frequencies, the proportion of low frequency energy and the low-frequency limit of the audio excerpt.
    \vspace*{-0.2cm}
    \item \textbf{roughness}: refers to the irregular and uneven sonic texture of a sound. It is estimated from the interaction of peaks and nearby bins within frequency spectral frames. When neighboring frequency components have peaks with similar amplitude, the sound is said to produce a ‘rough’ sensation.
    \vspace*{-0.2cm}
    \item \textbf{boominess}: refers to a sound with deep and loud resonant components.\footnote{\label{foot:notavailable}Description of the calculation method for this feature is not available to the authors at current time.}
    \vspace*{-0.2cm}
    \item \textbf{warmth}: refers to sounds that induce a sensation analogous to that caused by the physical temperature.$\:^\text{5}$ 
    \vspace*{-0.2cm}
    \item \textbf{sharpness}: refers to a sound that might cut if it were to take on physical form.$\:^\text{5}$ 
\end{itemize}

\subsection{Architecture Design and Training Procedure}\label{sec:arch}
In the following, we will introduce the architecture and training of DrumGAN, and will briefly describe the baseline model against which DrumGAN is evaluated.
\subsubsection{Generative Adversarial Networks}
\begin{figure}[t]
\centering
\includegraphics[scale=0.5, width=0.965\columnwidth]{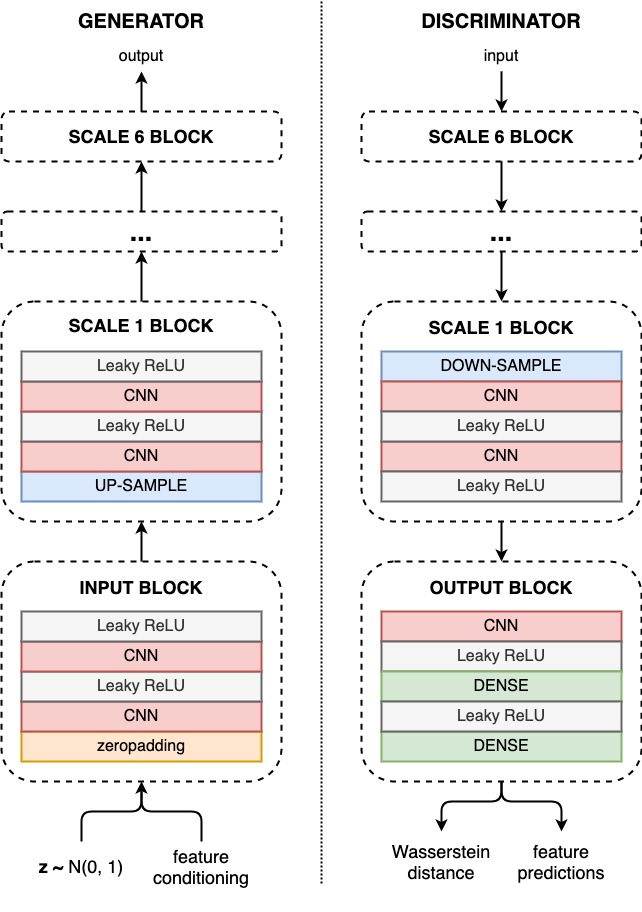}
\caption{Proposed architecture for DrumGAN (see Section \ref{sec:arch} for details).}
\label{fig:arch}

\end{figure}

Generative Adversarial Networks (GAN) are a family of training procedures inspired by game theory, in which a generative model competes against a discriminative adversary, that learns to distinguish whether a sample is real or fake \cite{Goodfellow2013}. The generative network, or Generator (G), estimates a distribution \(p_{g}\) over the data \(x\) by learning a mapping of an input noise \(p_{z}\) to data space as \(G_\theta(z)\), where \(G_\theta\) is a neural network implementing a differentiable function with parameters \(\theta\). Inversely, the discriminator \(D_\beta(x)\), with parameters $\beta$ is trained to output a single scalar indicating whether the input comes from the real data $p_r$ or from the generated distribution \(p_{g}\). Simultaneously, \(G\) is trained to produce samples that are identified as real by the discriminator. Competition drives both networks until an equilibrium point is reached and the generated examples are indistinguishable from the original data. For a Wasserstein GAN, as used in our experiments, the training criterion is formally defined as

\vspace{-0.5cm}

\begin{equation}
    \label{eq:gan}
    \begin{split}
        \min_{G} \max_{D} \Gamma(D,G) = \frac{1}{N} \sum_i D(x^{i}) - D(G(z^{i})).
    \end{split}
\end{equation}
\subsubsection{Proposed Architecture}
In the proposed architecture, the input to $G$ is a concatenation of the 7 conditioning features \(c\), described in Section \ref{sec:acommons}, and a random vector \(z\) with 128 components sampled from an independent Gaussian distribution. The resulting vector is fed through a stack of convolutional and box up-sampling blocks to generate the output signal \(x=G_\theta(z; c)\). In order to turn the 1D input vector into a 4D convolutional input, it is zero-padded in the time- and frequency-dimension (i.e., placed in the middle of the convolutional input with $128+7$ convolutional maps).  As depicted in Figure \ref{fig:arch}, the generator's input block performs this zero-padding followed by two convolutional layers with ReLU non-linearity. Each scale block is composed of one box up-sampling step at the input and two convolutional layers with filters of size $(3, 3)$. The number of feature maps decreases from low to high resolution as \{256, 128, 128, 128, 64, 32\}. Up-sampling of the temporal dimension is just performed after the 3rd scale block. We use a Leaky ReLU as activation functions and apply pixel normalization after every convolutional step (i.e., normalizing the norm over the output maps at each position). The discriminator \(D\) is composed of convolutional and down-sampling blocks,  mirroring $G$'s configuration. Given a batch of either real or generated STFT audio, \(D\) estimates the Wasserstein distance between the real and generated distributions \cite{Gulrajani2017}, and predicts the perceptual features accompanying the input audio in the case of a real batch, or those used for conditioning in the case of generated audio. In order to promote the usage of the conditioning information by $G$, we add an auxiliary Mean Squared Error (MSE) loss term to the objective function, following a similar approach as in \cite{OdenaOS17}. We use a gradient penalty of 10.0 to satisfy the Lipschitz continuity condition of Wasserstein GANs. The weights are initialized to zero and we apply layer-wise normalization at run-time using He's constant \cite{HeZRS15} to promote an equalized learning. A mini-batch standard deviation layer before the output block of \(D\) encourages \(G\) to generate more variety and, therefore, reduce mode collapse \cite{SalimansGZCRCC16}.

\subsubsection{Training Procedure}
Training follows the procedure of Progressive Growing of GANs (PGANs), first used for image generation \cite{Karras2017}, which has been successfully applied to audio synthesis of pitched sounds \cite{gansynth}. In a PGAN, the architecture is built dynamically during training. The process is divided into training iterations that progressively introduce new blocks to both the Generator and the Discriminator, as depicted in Figure \ref{fig:arch}.  While training, a blending parameter \(\alpha{}\) progressively fades in the gradient derived from the new blocks, minimizing possible  perturbation effects. The models are trained for 1.1M iterations on batches of 30, 30, 20, 20 12 and 12 samples, respectively for each scale. Each scale is trained during 200k iterations except the last one, which is trained up to 300k iterations. We employ Adam as the optimization method and a learning rate of 0.001 for both networks.

\subsubsection{The U-Net Baseline}\label{sec:unet}
As mentioned in the introduction, we compare DrumGAN against a previous work tackling the exact same task (i.e., neural synthesis of drums sounds, conditioned on the same perceptual features described in Section \ref{sec:acommons}), but using a U-Net architecture operating in the time domain \cite{aramires}. The U-Net model is trained to deterministically map the conditioning features (and an envelope of the same size as the output) to the output. The dataset used thereby consists of $~11$k drum samples obtained from Freesound\footnote{\scriptsize\url{www.freesound.org}}, which includes kicks, snares, cymbals, and other percussion sounds (referred to as \emph{Freesound drum subset} in the following).

\subsection{Evaluation}\label{sec:eval}
Assessing the quality of synthesized audio is hard to formalize making the evaluation of generative models for audio a challenging task. In the particular case of GANs, where no explicit likelihood maximization exists, a common evaluation approach is to measure the model's performance in a variety of surrogate tasks \cite{TheisOB15}. As described in the following, we evaluate our models against a diverse set of metrics that capture distinct aspects of the model's performance.

\subsubsection{Inception Score}\label{sec:inception}
The \emph{Inception Score (IS)} \cite{SalimansGZCRCC16} penalizes models that generate examples that are not classified into a single class with high confidence, as well as models whose examples belong to only a few of all the possible classes. It is defined as the mean KL divergence between the conditional class probabilities \(p(y|x)\), and the marginal distribution \(p(y)\) using the class predictions of an Inception classifier (see Eq. \ref{eq:iscore}). We train our Inception Net\footnote{\scriptsize\url{https://github.com/pytorch/vision/blob/master/torchvision/models/inception.py}} variant to classify kicks, snares and cymbals, from Mel-scaled magnitude STFT spectrograms using the same train/validation split of 90\% / 10\%, used throughout our experiments. As additional targets, we also train the model to predict the extracted perceptual features described in Section \ref{sec:acommons} (using mean-squared error cost).
    \begin{equation}
        IS = \exp{\big (E_x[KL(p(y|x)||p(y))]\big)}
        \label{eq:iscore}
    \end{equation}

    
\subsubsection{Fréchet Audio Distance (FAD)}
The \emph{Fréchet Audio Distance} compares the statistics of real and generated data computed from an embedding layer of a pre-trained VGG-like model\footnote{\scriptsize\url{https://github.com/google-research/google-research/tree/master/frechet\_audio\_distance}} \cite{fad}. FAD fits a continuous multivariate Gaussian to the output of the embedding layer for real and generated data and the distance between these is calculated as:
    \begin{equation}
        FAD = ||\mu{}_r - \mu{}_g||^2 + tr(\Sigma_r + \Sigma_g - 2\sqrt{\Sigma_r\Sigma_g})
        \label{eq:fad}
    \end{equation}
where \((\mu{}_r, \Sigma_r)\) and \((\mu{}_g, \Sigma{}_g)\) are the mean and co-variances of the embedding of real and generated data respectively. The lower the FAD, the smaller the distance between distributions of real and generated data. FAD is robust against noise, consistent with human judgments and more sensible to intra-class mode dropping than IS.

\subsubsection{Kernel Inception Distance (KID)}
The KID measures the dissimilarity between samples drawn independently from real and generated distributions \cite{BinkowskiSAG18}. It is defined as the squared Maximum Mean Discrepancy (MMD) between representations of the last layer of the Inception model (described in Section \ref{sec:inception}). A lower MMD means that the generated \(p_g\) and real \(p_r\) distributions are close to each other. We employ the unbiased estimator of the squared MMD \cite{GrettonBRSS12} between \(m\) samples \(x\thicksim p_r\) and \(n\) samples \(y\thicksim p_g\), for some fixed characteristic kernel function \(k\), defined as

\begin{footnotesize}
    \vspace{-0.3cm}
    \begin{equation}
            \begin{split}
                MMD^2 (X, Y) =
                    & \quad \frac{1}{m(m-1)}\sum_{i\neq j}^{m}k(x_i, x_j) \\ + & \quad \frac{1}{n(n-1)}\sum_{i\neq j}^{n}k(y_i, y_j) \\ - & \quad \frac{2}{mn}\sum_{i=1}^{m} \sum_{j=1}^{n} k(x_i,y_j).
            \end{split}
    \end{equation}
    \end{footnotesize}
    
Here, we use an inverse multi-quadratic kernel (IMQ) \(k(x,y) = 1 / (1+||x-y||^2 / 2 \gamma^2)\) with \(\gamma^2 = 8\) \cite{Rustamov}, which has a heavy tail and, hence, it is sensitive to outliers. We borrow this metric from the Computer Vision literature and apply it to the audio domain. We train a separate inception model on the FreeSound drum subset used for the U-Net baseline experiments (see Section \ref{sec:unet}). This is done to allow comparison of the inception-based metrics with DrumGAN. Since the FreeSound drum subset doesn't contain annotations of the instrument type, we train our variant on just the feature regression task, and restrict our comparison to KID and FAD, as these metrics do not compare class probabilities but embedding distributions.

\subsubsection{Feature Coherence}\label{sec:fcoherence}
We follow the methodology proposed by \cite{aramires} for evaluating the feature control coherence. The goal is to assess whether increasing or decreasing a specific feature value of the conditioning input yields the corresponding change of that feature in the synthesized audio. To this end, a specific feature $i$ is set to $0.2$ (low), $0.5$ (mid), and $0.8$ (high), keeping the other features and the input noise fixed. The resulting outputs $x_{\text{low}}^i$, $x_{\text{mid}}^i$, $x_{\text{high}}^i$ are then evaluated with the Audio Commons Timbre Models (yielding features $fx^i$). Then, it is assessed if the feature of interest changed as expected (i.e., $fx_{\text{low}}^i < fx_{\text{mid}}^i < fx_{\text{high}}^i$). More precisely, three conditions are evaluated: E1: $fx_{\text{low}}^i < fx_{\text{high}}^i$, E2: $fx_{\text{mid}}^i < fx_{\text{high}}^i$, and E3: $fx_{\text{low}}^i < fx_{\text{mid}}^i$. We perform these three tests $1000$ times for each feature, always with different random input noise and different configurations of the other features (sampled from the evaluation set). The resulting accuracies are reported.

\vspace*{-2mm}
\section{Results and Discussion}\label{sec:res}
In this section, we briefly describe our subjective impression when listening to the model output, and we will give an extended discussion on the quantitative analysis, including the comparison with the baseline U-Net architecture.
\vspace*{-3mm}
\subsection{Subjective Evaluation and Generation Tests}\label{sec:qualitative}
\vspace*{-1mm}
The results of the qualitative experiments discussed in this section can be found on the accompaniment website.\footnote{\scriptsize\url{https://sites.google.com/view/drumgan}} In general, conditional DrumGAN seems to have better quality than its unconditional counterpart and substantially better than the U-Net baseline (see Section \ref{sec:unet}). In the absence of more reliable baselines, we argue that the perceived quality of DrumGAN is comparable to that of previous state-of-the-art work on adversarial audio synthesis of drums \cite{wavegan}.

We also perform radial and spherical interpolation experiments (with respect to the Gaussian prior) between random points selected in the latent space of DrumGAN. Both interpolations yield smooth and perceptually linear transitions in the audio domain. We notice that radial interpolation tend to change the percussion type (i.e., kick, snare, cymbal) of the output, while spherical interpolation affects other properties (like within-class timbral characteristics and envelope) of the synthesized audio. This gives a hint on how the latent manifold is structured.


\subsection{Quantitative Results}
\begin{table}
 \begin{center}
 \begin{tabular}{lllll}
  \toprule
   & IS & KID & FAD\\
  \midrule
  \emph{real data}  & 2.26 & 0.05 & 0.00\\
  \emph{train feats}  & \textbf{2.19} & 0.39 & 0.77\\
  \emph{val feats}  & 2.18 & \textbf{0.35} & 0.76\\
  \emph{rand feats} & 2.09 &  1.36 & \textbf{0.70}\\
  \emph{unconditional} & \textbf{2.19} &  1.07 & 1.00\\
  \bottomrule
 \end{tabular}
\end{center}
\vspace{-2mm}
 \caption{Results of Inception Score (IS, higher is better), Kernel Inception Distance (KID, lower is better) and Fréchet Audio Distance (FAD, lower is better), scored by DrumGAN under different conditioning settings, against real data and the unconditional baseline. The metrics are computed over $50$k samples, except for \emph{val feats}, where $30$k samples are used (i.e., the validation set size).}
 \label{tab:eval_pgan}
\end{table}

\begin{table}
 \begin{center}
 \begin{tabular}{llll}
  \toprule
   & KID & FAD\\
  \midrule
  \emph{real data} &  0.04 & 0.00\\
  \emph{real feats}  & 1.45 & 3.09\\
       \emph{rand feats} &  13.94 &  3.17\\
  \bottomrule
 \end{tabular}
\end{center}
\vspace{-2mm}
 \caption{Results of Kernel Inception Distance (KID) and Fréchet Audio Distance (FAD), scored by the U-Net baseline \cite{aramires} when conditioning the model on feature configurations from the real data and on randomly sampled features. The metrics are computed over $11$k samples (i.e., the Freesound drum subset size).}
 \label{tab:eval_unet}
 \vspace{-5mm}
\end{table}

\subsubsection{Scores and Distances}
Table \ref{tab:eval_pgan} shows the DrumGAN results for the Inception Score (IS), the Kernel Inception Distance (KID), and the Fréchet Audio Distance (FAD), as described in Section \ref{sec:eval}. These metrics are calculated on the synthesized drum sounds of the model, based on different conditioning settings. Besides the unconditional setting of DrumGAN (\emph{unconditional}), we use feature configurations from the train set (\emph{train feats}), the valid set (\emph{valid feats}), and features randomly sampled from a uniform distribution (\emph{rand feats}).
The IS of DrumGAN samples is close to that of the real data in most settings. This means that the model outputs are clearly assignable to either of the respective percussion-type classes (i.e., low entropy for kick, snare, and cymbal posteriors), and that it doesn't omit any of them (i.e., high entropy for the marginal over all classes). The IS is slightly reduced for random conditioning features, indicating that using uncommon conditioning configurations makes the outputs more ambiguous with respect to specific percussion types. While FAD is a measure for the perceived quality of the individual sounds (measuring co-variances within data instances), the KID reflects if the generated data overall follows the distribution of the real data. Therefore, it is interesting to see that \emph{rand feats} cause outputs which overall do not follow the distribution of the real data (i.e., high KID), but the individual outputs are still plausible percussion samples (i.e., low FAD). This quantitative result is in-line with the perceived quality of the generated samples (see Section \ref{sec:qualitative}). In the unconditional setting, both KID and FAD are worse, indicating that feature conditioning helps the model to both generate data following the true distribution, overall, as well as in individual samples.

Table \ref{tab:eval_unet} shows the evaluation results for the U-Net architecture (see Section \ref{sec:unet}). As the train / valid split for the Freesound drum subset (on which the U-Net was trained) is not available to the authors, the U-Net model is tested using the features of the full Freesound drum subset (\emph{real feats}), as well as random features. Also, we do not report the IS for the U-Net architecture, as it was trained on data without percussion-type labels, making it impossible to train the inception model on such targets. As a baseline, all metrics are also evaluated on the real data on which the respective models were trained. While evaluation on the real data is straight-forward for the IS (i.e., just using the original data instead of the generated data to obtain the statistics), both KID and FAD are measures usually \emph{comparing} the statistics between features of real and generated data. Therefore, for the \emph{real data} baseline, we split the real data into two equal parts and compare those with each other in order to obtain KID and FAD.
The performance of the U-Net approach on both, KID and FAD is considerably worse than that of DrumGAN. While the KID for \emph{real feats} is still comparable to that of DrumGAN (indicating a distribution similar to that of the real data), the high FAD indicates that the generated samples are not perceptually similar to the real samples. When using random feature combinations this trend is accentuated moderately in the case of FAD, and particularly in the case of the KID, reaching a maximum of almost $14$. This is, however, intelligible, as the output of the U-Net depends only on the input features in a deterministic way. Therefore, it is to expect that the distribution over output samples changes fully when fully changing the distribution of the inputs.

\vspace*{-1mm}
\subsubsection{Feature Coherence}
\begin{table}
 \small
 \begin{center}
 \begin{tabular}{llllclll}
 \toprule
  & \multicolumn{3}{c}{U-Net} & & \multicolumn{3}{c}{DrumGAN} \\
  \cmidrule{2-4} \cmidrule{6-8}
  Feature & E1 & E2 & E3 && E1 & E2 & E3\\
  \midrule
  brightness  & \textbf{0.99} & \textbf{0.99} & \textbf{1.00} && 0.74 & 0.71 & 0.70\\
  hardness  & 0.64 & \textbf{0.65} & 0.59 && 0.64 &  0.64 & \textbf{0.62}\\
  depth  & \textbf{0.94} & 0.65 & \textbf{0.94} && 0.79 & \textbf{0.72} & 0.74\\
  roughness  & 0.63 & 0.59 & 0.57 && \textbf{0.72} &  \textbf{0.68} & \textbf{0.67}\\
   boominess & \textbf{0.98} & \textbf{0.82} & \textbf{0.98} && 0.80 & 0.74 & 0.77\\
   warmth & \textbf{0.92} & \textbf{0.79} & \textbf{0.91} && 0.76 & 0.71 & 0.71\\
    sharpness & 0.63 & 0.77 & 0.45 && \textbf{0.84} & \textbf{0.82} & \textbf{0.82} \\
\midrule
average & \textbf{0.83} & \textbf{0.76} & \textbf{0.78} && 0.76 & 0.72 & 0.72 \\
  \bottomrule
 \end{tabular}
\end{center}
 \vspace*{-3mm}
 \caption{Mean accuracies for the feature coherence tests on samples generated with the baseline U-Net \cite{aramires} and DrumGAN.}
 \vspace*{-4mm}
 \label{tab:feat_coherence}
\end{table}

Table \ref{tab:feat_coherence} shows the accuracy of the three feature coherence tests explained in Section \ref{sec:fcoherence}. Note that, as both models were trained on different data, the figures of the two models are not directly comparable. However, also reporting the figures of the U-Net approach should provide some context on the performance of our proposed model. In addition, as both works use the same feature extractors and claim that the conditional features are used to shape the same characteristics of the output, we consider the figures from the U-Net approach a useful reference. We can see that for about half the features, the U-Net approach reaches close to $100\%$ accuracy. Referring to the descriptions on how the features are computed it seems that the U-Net approach reaches particularly high accuracies for features which are computed by looking at the global frequency distribution of the audio sample, taking into account spectral centroid and relations between high and low frequencies (e.g., brightness and depth). U-Net performs considerably worse for features which take into account the temporal evolution of the sound (e.g., hardness) or more complex relationships between frequencies (e.g., roughness). While DrumGAN performs worse on average on these tests, the results seem to be more consistent, with less very high, but also less rather low accuracy values (note that the random-guessing baseline is $0.5$ for all the tests). The reason for not performing better on average may lie in the fact that DrumGAN is trained in an adversarial fashion, where the dataset distribution is enforced, in addition to obeying the conditioned characteristics. In contrast, in the U-Net approach the model is trained deterministically to map the conditioning features to the output, which makes it easier to satisfy the simpler characteristics, like generating a lot of low- or high-frequency content. However, this deterministic mapping results in a lower audio quality and a worse approximation to the true data distribution, as it can be seen in the KID and FAD figures, described above.

\vspace{-0.3cm}
\section{Conclusions and Future Work} \label{sec:con}
\vspace{-0.1cm}
In this work, we performed percussive sound synthesis using a GAN architecture that enables steering the synthesis process using musically meaningful controls. To this end, we collected a dataset of approximately 300k audio samples containing kicks, snares, and cymbals. We extracted a set of timbral features, describing high-level semantics of the sound, and used these as conditional input to our model. We encouraged the generator to use the conditioning information by performing an auxiliary feature regression task in the discriminator and adding the corresponding MSE loss term to the objective function. In order to assess whether the feature conditioning improves the generative process, we trained a model in a completely unsupervised manner for comparison. We evaluated the models by comparing various metrics, each reflecting different characteristics of the generation process. Additionally, we compared the coherence of the feature control against previous work. Results showed that DrumGAN generates high-quality drum samples and provides meaningful control over the audio generation. The conditioning information was proven useful and enabled the network to better approximate the real distribution of the data. As future work, we will focus on scaling the model to work with audio production standards (e.g., 44.1kHz sample rate, stereo audio), and implement a plugin that can be used in a conventional Digital Audio Workstation (DAW).

\section{Acknowledgements}
This research was supported by the European Union's Horizon 2020  research and innovation program under the Marie Skłodowska-Curie grant agreement No. 765068 (MIP-Frontiers).
\bibliography{ISMIRtemplate}

%
%
%
%

\end{document}